# Fourfold Symmetry of Anisotropic Magnetoresistance in Epitaxial $Fe_3O_4$ Thin Films

C.R.Hu, J.Zhu, G. Chen, J.X.Li and Y.Z.Wu[*]

Department of Physics, State Key Laboratory of Surface Physics, and Advanced Materials Laboratory, Fudan University, Shanghai 200433, P. R. China

**ABSTRACT**

We studied the angular dependence of anisotropic magnetoresistance (AMR) and Planar Hall effect (PHE) at various temperatures in high quality epitaxial $Fe_3O_4$ films grown on MgO(001) substrates. The PHE contains only a twofold angular dependence, but the AMR below 200K is constituted with both twofold and fourfold symmetric terms. A quantitative fitting based on a phenomenological model indicates the nonmonotonics temperature dependence of the twofold component of AMR can be ascribed to the competition between the $\cos^2\theta$ term and the $\cos^4\theta$ term. A unidirectional component was observed in the angular dependent AMR. The fourfold symmetric AMR also existed for the magnetic field rotating in the plane perpendicular to the current. Our results indicate the AMR and PHE in single crystalline films have different origins, and also prove that the origin of the four-fold symmetry of AMR is related to the lattice symmetry rather than the spin scattering near the antiphase boundaries.

**INTRODUCTION**

Anisotropic magnetoresistance (AMR) in ferromagnets depends on the orientation of the magnetization with respect to the electric current direction in the material, and is believed to be a consequence of spin-orbit coupling [1,2,3,4]. Spin-orbit coupling can induce the mixing of spin-up and spin-down states, and such mixing depends on the magnetization direction, thus giving rise to a magnetization-direction dependent scattering rate. Although the AMR effect has shown great applications in magnetic recording technologies, the understanding of the microscopic physics of this spin-orbit coupling induced effect is still relatively poor.

For polycrystalline ferromagnetic materials, in terms of the single domain model, with the magnetization lying in the plane, the in-plane longitudinal and transversal resistivity can be respectively expressed as [1,2,3]:

$$\rho_{xx} = \rho_\perp + (\rho_{//} - \rho_\perp)\cos^2\theta, \quad (1)$$

$$\rho_{xy} = (\rho_{//} - \rho_\perp)\sin\theta\cos\theta, \quad (2)$$

Where $\rho_\perp$ and $\rho_{//}$ are the resistivity for the magnetization perpendicular and parallel to the current respectively, and θ is the angle between current and magnetization. Eqn. (1) and (2) are conventionally used to describe the AMR and planar Hall effect (PHE), so both AMR and PHE have a twofold symmetry about the current direction with the rotation of M and are believed to relate to each other. Eqn. (1) and (2) are independent of the crystal axis, so they should be adapted only for the polycrystalline samples, but they have also been used to describe the AMR and PHE effect in epitaxial films [5,6,7]. However, in single crystal or epitaxial thin films, the AMR effect should reflect the crystalline symmetry due to its origin from spin-orbit coupling [8]. Recently, the deviation of AMR away from $\cos^2\theta$ has been observed in manganites [9,10,11], (Ga,Mn)As[12,13], as well as in $Fe_3O_4$ films [14,15,16,17]. A fourfold oscillation was observed superimposed on the two-fold oscillation in the longitudinal resistivity, which is usually recognized from high rank resistivity tensors, but the microscopic physical origin of this high order symmetric AMR and the evolution from twofold symmetry to fourfold symmetry are not yet well understood.

Magnetite ($Fe_3O_4$) is an important 3d transition-metal oxide due to its half metallic properties, so it is considered as a potential material for spintronics applications [18,19,20], and the magnetoresistance properties of $Fe_3O_4$ films have attracted great interest [21,22,23,24]. $Fe_3O_4$ films were found to contain a giant PHE [25,26], which is proposed to be used for designing a PHE-based magnetic random access memory (MRAM). Recently, an additional fourfold symmetry of the AMR effect in $Fe_3O_4$ films was first observed in longitudinal resistivity measurement geometry by Ramos et .al [14], however it has not been studied whether PHE also has the similar anomalous behaviors. The twofold and fourfold contributions of AMR have been separated by a phenomenological equation [14], and the twofold component shows nonmonotonic temperature dependence with a minimum at ~150K. However, the origin of this nonmonotonic temperature dependence is still unclear. Moreover, Ramos et al [14] attributed the fourfold symmetry of AMR to the consequence of spin-orbit enhancement induced by the change-order formation. But Li et al. [16,17] proposed magnetocrystalline anisotropy and antiphase boundaries (APBs) as the origin of the fourfold symmetric AMR. Identifying the origin of this fourfold symmetric AMR effect is required for further understanding the mechanism of the magnetoresistance effect in magnetite.

In this paper, we report our study on the temperature dependence of AMR in epitaxial $Fe_3O_4$ films. The simultaneous study of AMR and PHE clearly show their different origins in single crystalline films. The quantitative fitting based on a phenomenological model indicates the nonmonotonic temperature dependence of the twofold component AMR can be ascribed to the competition between $\cos^2\theta$ and $\cos^4\theta$ terms. A unidirectional component was also clearly observed in the angular dependent AMR. The origin of the fourfold symmetric AMR was proven to be related to the lattice symmetry rather than the spin scattering near the APBs.

**EXPERIMENTAL**

$Fe_3O_4$ thin films were grown on MgO(001) single-crystal substrates in an ultra-high vacuum chamber with a base pressure of $2\times10^{-10}$ Torr. MgO(001) substrates with a miscut angle less than $0.5^o$ were first annealed at 600 °C in UHV for 1hour, then a 10nm MgO seed layer was grown at 500 ℃ by pulsed laser deposition (PLD) from a MgO target. The reflection high energy electron diffraction (RHEED) patterns indicate the surface morphology was improved by the growth of a MgO seed layer. $Fe_3O_4$ thin films were epitaxied on the MgO substrate by evaporating the Fe atoms at an oxygen pressure of ~$1\times10^{-5}$ Torr. The substrate temperature during the growth was 250 ℃. The growth rate was determined by a quartz crystal thickness monitor. The epitaxial growth of $Fe_3O_4$ films could be confirmed by the sharp RHEED patterns shown in Fig.1(a) and (b).

The high quality epitaxial growth of $Fe_3O_4$ films could also be proved by the X-ray diffraction data in Fig.1(c). The X-ray diffraction data were obtained at beamline BL14B1 of the Shanghai Synchrotron Radiation Facility (SSRF) at a wavelength of 1.2398 Å. BL14B1 is a beamline based on bending magnet, and a Si(111) double crystal monochromator was employed to monochromatize the beam. A NaI scintillation detector was used for data collection. The symmetrical θ/2θ scan around the MgO(002) peak shows the $Fe_3O_4$(004) reflection peak, which gives the expected orientation $Fe_3O_4$ [001]//MgO [001]. The out-of-plane lattice parameter of the $Fe_3O_4$ film extracted from the peak position is $a_{out}$~8.40Å, almost the same as the bulk $Fe_3O_4$ lattice parameter. It should be noted that we were able to observe Laue oscillations up to the 15th order, which is normally taken as an indication of a very high crystalline coherence.

Resistance versus temperature measurements were performed using a standard four probe method on a patterned sample. Prior to the transport measurements, the samples were patterned into the Hall geometry by two-step lithography. The optical lithography process included the following processes: photoresist spin coating, mask alignment and illumination, photoresist development, Ar+ ion beam etching, Au pad fabrication and wire linking onto a sample puck of Quantum Design physical property measurement system (PPMS) equipped with a sample rotator. The schematic diagram of the sample after the optical lithography is shown in Fig. 2(a). The current was kept 5μA flowing along the [100] direction in all the transport measurements. In this work, the AMR is defined as $AMR = \Delta\rho / \rho_{min}(\%) = [(\rho(\theta) - \rho_{min})/\rho_{min}] \times 100$ [14], where θ is the angle between the magnetic field and the current, and the $\rho_{min}$ is the minimum resistivity for each scan.

**RESULTS AND DISCUSSION**

Fig.2(c) shows the relative amplitude of the angular dependent magnetoresistance for the 46.8nm $Fe_3O_4$ with the 1.5T applied field at difference temperature. Fig. 2(b) shows a typical hysteresis loop measured by Magneto-Optic Kerr Effect (MOKE) for the magnetic field along the <100> direction at room temperature. The hysteresis loop shows the saturation field is less than 0.15T, so the total magnetization can be considered to follow the field direction during the transport measurement with the 1.5T in-plane applied field. For temperature higher than 200K, the angular dependence of AMR shows only a two-fold symmetry, i.e. with the peaks at $0^o$ and $180^o$, and the valleys at $90^o$ and $270^o$. When the sample was cooled below 200K, an additional four-fold symmetry appeared with peaks at $90^o$ and $270^o$. In fact, such a four-fold symmetry may also exist at high temperature, since the AMR curve at 300K cannot be well fitted by a $\cos^2\theta$ function. Our result is consistent with the results in Refs [14,15,16,17], and such a high order symmetry of AMR has also been observed in Manganites [9,10,11] and in diluted magnetic semiconductors [12,13].

From Eqn. (1) and (2), the same physical origin is expected for AMR and PHE in a ferromagnet. If so, the four-fold symmetry could be expected in the angular dependent PHE. Fig.3 shows the angular dependent PHE measurement on the same sample at different temperatures. All the PHE curves contain only twofold symmetry. The PHE effect contains one clear sign reversal at ~130K. We also observed a giant PHE at low temperature, and the value of the PHE is about two orders larger than the value at room temperature, consistent with a previous report [6]. Nevertheless, the existence of only two-fold symmetry of PHE at low temperature indicates the different physical origin of AMR and PHE in single crystalline films.

A phenomenological model has been widely applied to understand the origin of the different symmetry in AMR and PHE in ferromagnetic crystals [1-3]. The resistivity tensor $\rho_{ij}$ depends on the direction cosines, $\alpha_i$, of the magnetization, and may be expressed as series expansions in ascending power of $\alpha_i$

$$\rho_{ij} = a_{ij} + a_{kij}\alpha_k + a_{klij}\alpha_k\alpha_l + a_{klmij}\alpha_k\alpha_l\alpha_m + a_{klmnij}\alpha_k\alpha_l\alpha_m\alpha_n + \ldots \quad (3)$$

Here, $\alpha_{ij}$, $\alpha_{kij}$, $\alpha_{klij}$, $\alpha_{klmij}$ and $\alpha_{klmnij}$ are the elements of the resistivity tensor of various orders, and $i$ and $j$ can be in any of the three orthogonal directions. If the magnetization lies strictly in the film plane, the resistivities $\rho_{xx}$ and $\rho_{xy}$ are related to the AMR and PHE, respectively.

Many of the matrix elements actually vanish due to the symmetry of the cubic crystal and the Onsager relation [27]. For a (001) cubic crystalline film with in-plane magnetization and the current flowing along <100> directions, the $\alpha_i$s are given by $\alpha_1 = \cos\theta$, $\alpha_2 = \sin\theta$ and $\alpha_3 = 0$, then AMR and PHE can be expressed as [2,14,28]:

$$\rho_{xx} = C_0 + C_1\cos^2\theta + C_2\cos^4\theta \quad (4)$$

$$\rho_{xy} = C_4\sin\theta\cos\theta \quad (5)$$

With $C_0 = a_{11} + a_{1122} + a_{111122}$, $C_1 = a_{1111} - a_{1122} - 2a_{111122} + a_{112211}$, $C_2 = a_{111111} + a_{111122} - a_{112211}$, $C_4 = a_{2323} + a_{111212}$. So, as a direct consequence of the crystal symmetry, AMR contains both twofold and fourfold symmetry terms, whereas PHE contains only twofold symmetry, which agrees well with our experiments. Eqn. (4) is based on the power expansions in terms of $\cos^n\theta$, but can also be modified with the symmetry point of view as follows [14]:

$$\rho_{xx} = A_0 + A_u\cos 2\theta + A_c\cos 4\theta \quad (6)$$

where $A_u = (C_1 + C_2)/2$ and $A_c = C_2/8$. Since Eqn. (4) and (5) are the original formulas derived directly from the phenomenological model in Eqn. (3), here we tried to use Eqn. (4) and (5) to fit the angular dependence of AMR and PHE, and to obtain the temperature dependence of the coefficients $C_1$, $C_2$ and $C_4$, which may provide a better understanding of the angular dependent magnetoresistance effect.

However, in Fig.2, the AMR of $Fe_3O_4$ film contains one obvious unidirectional component,

i.e. $\rho(0^o) \neq \rho(180^o)$, and such a unidirectional component of AMR can also be found in previous literature [14,15,16], but has never been discussed. As shown in Fig.4, the fitting by using Eqn. (4) and (5) does not agree with the experimental data. So in order to better fit the angular dependence of AMR and PHE, we add one additional unidirectional term into Eqn. (4) and (5) as follows:

$$\rho_{xx} = C_0 + C_U^{AMR} \cos(\theta - \theta_U^{AMR}) + C_1 \cos^2 \theta + C_2 \cos^4 \theta \qquad (7)$$

$$\rho_{xy} = C_U^{PHE} \cos(\theta - \theta_U^{PHE}) + C_4 \sin \theta \cos \theta \qquad (8)$$

Here $C_U^{AMR}$ and $C_U^{PHE}$ are the coefficients of the unidirectional terms and $\theta_U^{AMR}$ and $\theta_U^{PHE}$ are the angles of the easy axis direction. The fittings using Eqn. (7) and (8) agree very well with the experimental data, as shown in Fig. 4. We found that the easy axis of the unidirectional term is along the current for AMR, but perpendicular to the current for PHE. Then we fixed $\theta_U^{AMR} = 180^o$ and $\theta_U^{PHE} = -90^o$, and fitted the angular dependence of AMR and PHE using Eqn.(7) and (8). The fitted coefficients as a function of temperature are plotted in Fig.5. The coefficients $C_1$ and $C_4$ increase monotonically from negative to positive with increasing temperature and cross zero at a certain temperature, but $C_2$ exhibits a different trend, decreasing monotonically with the temperature. Such different temperature dependencies may not be surprising if one considers the fact that $C_2$ only contains the sixth rank tensors and both $C_1$ and $C_4$ include the fourth rank tensors and sixth rank tensors, so our results indicate that the fourth rank tensors and the sixth rank tensors may have different temperature dependencies. Ramos et. al [14] used Eqn.(8) to obtain the temperature dependent uniaxial component $A_u$, which has a nonmonotonic change and clearly shows a minimum at ~160K. The similar temperature dependence can be obtained if we add $C_1$ and $C_2$ together, as shown in the inset of Fig.5(a), so the nonmonotonic behavior of $A_u$ is just the result of the competition between $C_1$ and $C_2$.

The unidirectional terms $C_U^{AMR}$ and $C_U^{PHE}$ also show a monotonic change with temperature, and they are one order smaller than other coefficients. Although such unidirectional behaviors can also be seen in previous reports [14,15,16], the origin of such unidirectional terms is still not clear, and this behavior is not expected based on the symmetry analysis, as shown in Eqn.(4) and (5), without the broken of reversal symmetry. Muduli et al [28] also discovered an antisymmetric contribution of PHE in $Fe_3Si$ film grown on a GaAs(113) vicinal surface, which is caused by the second hall effect due to the miscut angle of the substrate. The small miscut angle of the MgO substrate is hard to be avoided, although the vicinal angle of the MgO(001) substrate can be guaranteed to be smaller than $0.5^o$, it is possible that such a small vicinal angle can induce a tiny unidirectional contribution in the angular dependence of AMR and PHE.

We will next discuss the origin of the fourfold symmetric AMR effect in $Fe_3O_4$ epitaxial films. The fourfold symmetric AMR in manganite films was attributed to the magnetocrystalline anisotropy [10]. When the external magnetic field is strong enough to overcome the magnetocrystalline anisotropy field, the spin distribution should be completely determined by the field direction, then the fourfold symmetric AMR can be suppressed by the higher magnetic field and degenerate into twofold symmetry [11]. Such magnetocrystalline anisotropy was also proposed to be the origin of the fourfold symmetry in $Fe_3O_4$ films [16,17]. However, this explanation cannot agree with the fact that the fourfold symmetric AMR effect in $Fe_3O_4$ films increases with the magnetic field up to 14T [14].

Another proposed origin of the fourfold symmetric AMR is the spin dependent scattering near the APBs. The APBs in $Fe_3O_4$ film are formed as a natural growth defect due to the differences in the translational and rotational symmetry between $Fe_3O_4$ and MgO [21,22,23,29,30]. It is usually believed that the APBs in $Fe_3O_4$ film will enhance the magnetoresistance due to the additional spin scattering induced by APBs [22,23,26,29]. In fact, the fourfold symmetry doesn't exist in a single APB since each APB only contains twofold symmetry. However, the distribution of the APBs' in-plane orientation has a fourfold symmetry, with the dominant occupation along the <100> direction [30], which may result in an overall fourfold spin scattering and induce the fourfold symmetric AMR. On the other hand, APBs are not expected to exist in the single crystalline $Fe_3O_4$, but Ramos et. al also observed the fourfold symmetric AMR in single crystalline $Fe_3O_4$ with a similar amplitude as in the $Fe_3O_4$ film [14], so this experimental result indicates that the fourfold symmetric AMR is not necessarily related to the APBs in the film. In order to identify the correlation between the APBs and the fourfold symmetric AMR, it is required to do the measurement on the APBs with only twofold symmetry, then to check whether the fourfold symmetry AMR can still exist. It should be noted that the APBs usually extend through the film, then they only contain the twofold symmetry in the cross-section plane, as shown by high resolution transmission electron microscopy (HRTEM) images [31]. We performed an additional magnetoresistance measurement in which the magnetic field rotated in the plane perpendicular to the current, as shown in Fig. 6(a). At high temperature, AMR only presents weak twofold symmetry, which is not surprising due to the symmetry broken between in-plane direction and out-of-plane direction. At low temperature, the fourfold symmetry AMR clearly appears, as shown in Fig. 6(b), and such fourfold symmetry appearing at low temperature also increases with the increasing applied magnetic field, so our results can rule out the possible origin related to the APBs since the APBs only contain twofold symmetry in the y-z plane. If the results in Fig.2 and Fig.6 are compared, it is clear that the AMR amplitude is quite similar in both measurements. The angular dependent AMR shows the peaks for the magnetization along <100> directions and the valleys along <110> directions, indicating the fourfold symmetry of AMR is related to the cubic crystalline structure. Generally speaking, the AMR should reflect the symmetry of the electron band structure that is associated with the lattice symmetry, since it is induced by spin-orbit coupling. For magnetic oxide films, the rotation of the magnetization may deform the local electron orbits through the spin-orbit interaction, which can change the overlapping of the orbits of the neighboring ions and further modulate the conductivity [9]. Such orbit deformation through spin-orbit interaction obviously should follow the symmetry of the lattice, and may also have strong dependence on the applied magnetic field. Ramos et. al also interpreted that the fourfold symmetric AMR is due to the formation of charge ordering, which also follows the cubic structural symmetry [14]. It remains an open question how to link the anisotropic magnetoresistance and the lattice symmetry, which requires further investigation both experimentally and theoretically.

**CONCLUSION**

In summary, the angular dependent AMR and PHE of epitaxial $Fe_3O_4$ films have been studied at various temperatures. The PHE only contains a twofold angular dependence, but the AMR below 200K is constituted by not only a twofold symmetric term but also a fourfold symmetric term. This result can be well explained by a phenomenological model. Our measurement performed for the magnetic field rotating in the plane perpendicular to the current proved that the high order symmetric AMR at low temperature is due to the symmetry of the lattice, and is not related to the spin scattering near the APBs in the $Fe_3O_4$ film. The results obtained can contribute to the understanding of high-order AMR in magnetic materials.

ACKNOWLEDGMENTS

The authors thank beamline BL14B1 (Shanghai Synchrotron Radiation Facility) for providing the beam time. This work was supported by NSFC and MOST (Nos. 2011CB921801, 2009CB929203 and 2010DFA52220) of China, SHEDF, STCSM, and the Fok Ying Tong education foundation.

Figures:

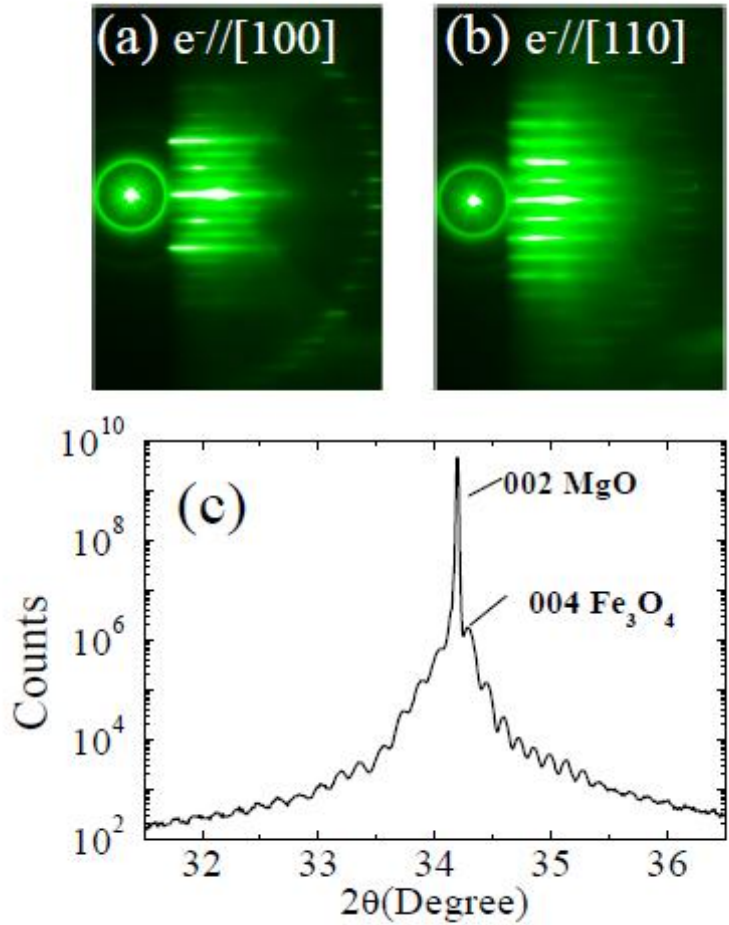

Fig. 1(a) and (b) The RHEED patterns with the electron along the <100> and <110> directions from a 41.8nm thick $Fe_3O_4$ film grown on MgO (001). (c) X-ray diffraction diagram (θ/2θ scan) around the (002) peak of the MgO substrate.

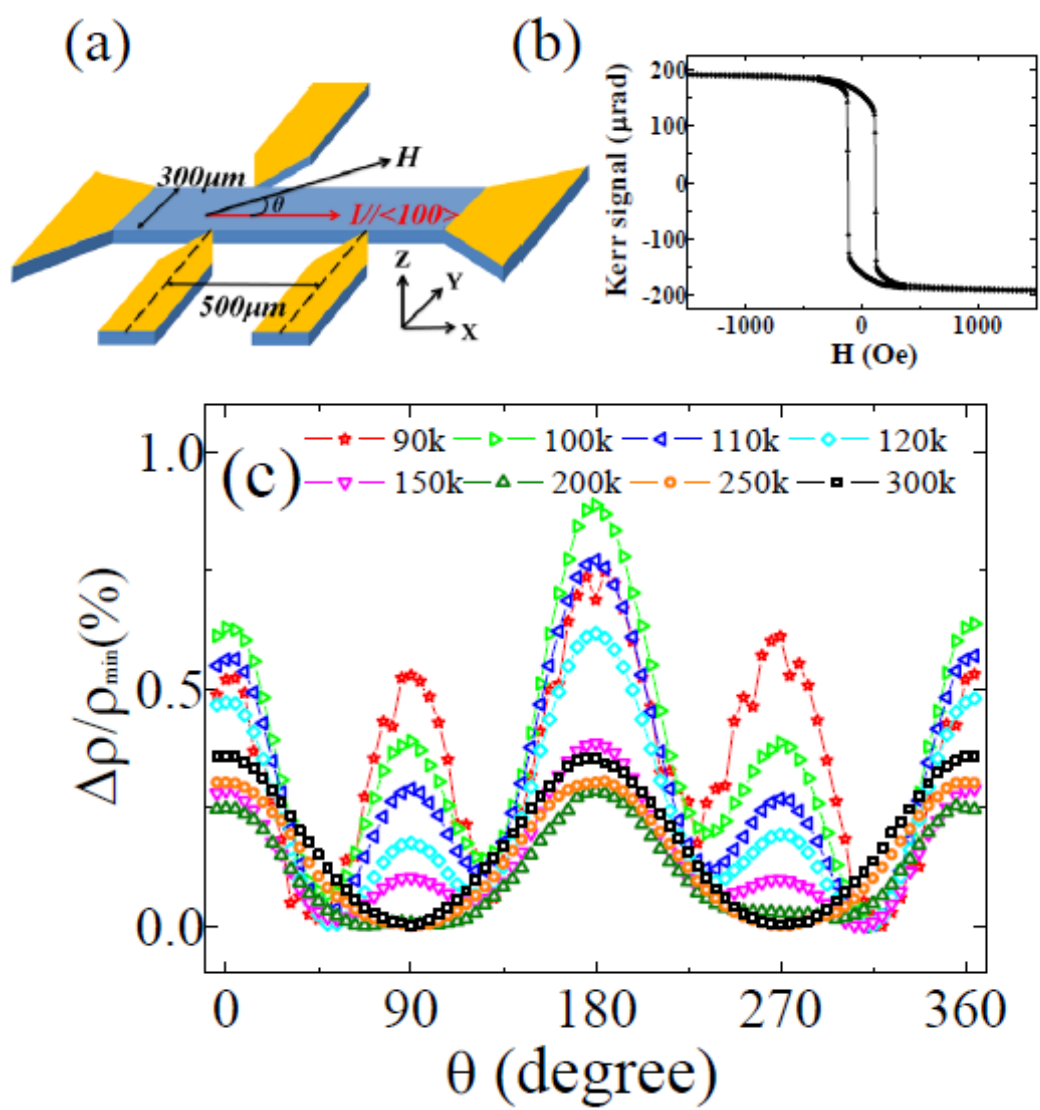

Fig.2. (a) Schematic diagram of the sample after optical lithography. (b) A typical hysteresis loop measured by MOKE. (c) The angular dependent magnetoresistance at various temperatures from a 46.8nm $Fe_3O_4$ film with a 1.5T applied field.

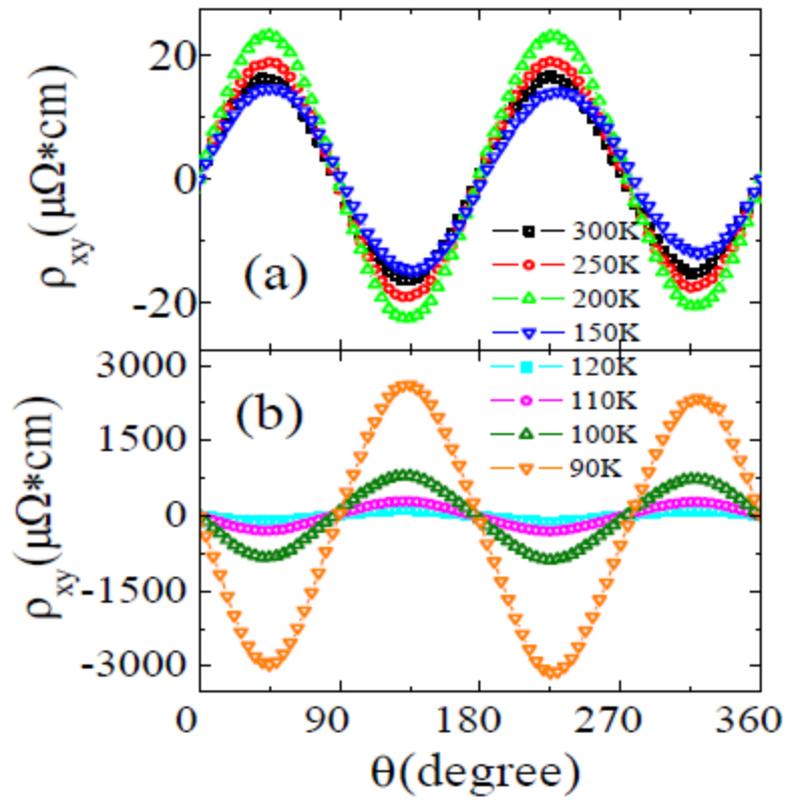

Fig.3.The in-plane angular dependence PHE resistivity $\rho_{xy}$ of the 46.8nm $Fe_3O_4$ film at different temperatures with a 1.5T applied field.

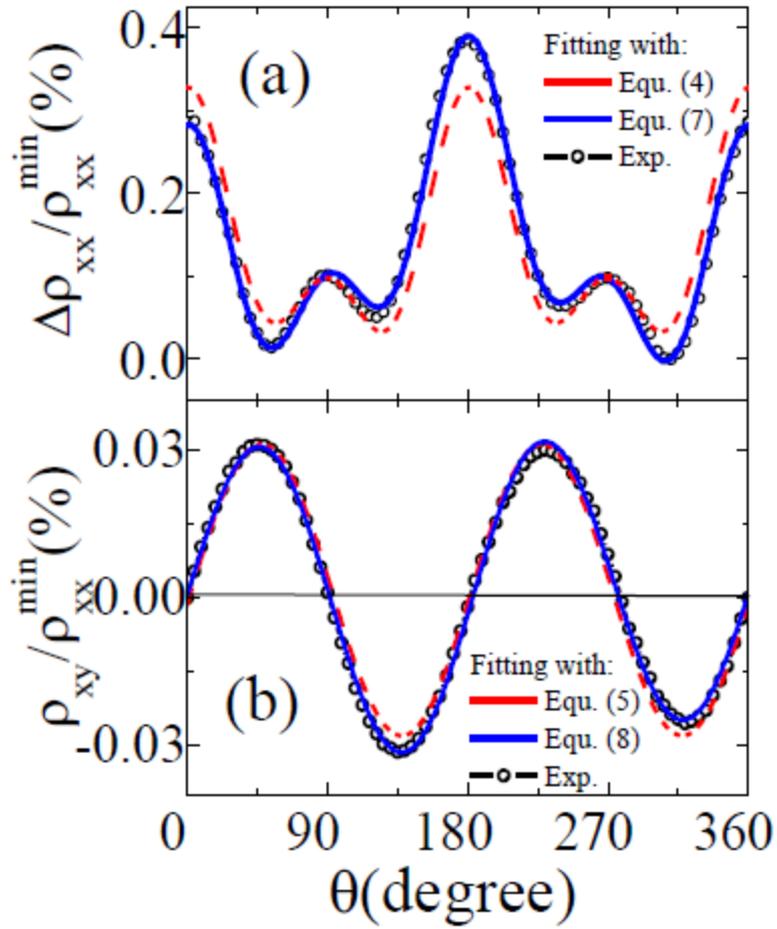

Fig. 4.(a) AMR and (b) PHE of the 46.8nm $Fe_3O_4$ film as a function of in-plane field orientations measured at 150K with a 1.5T applied field. The red dashed lines are the fitting curves with Eqn. (4) and (5), and the blue solid lines are the fitting curves with Eqn. (7) and (8). This result indicates that a unidirectional term has to be included in the fitting.

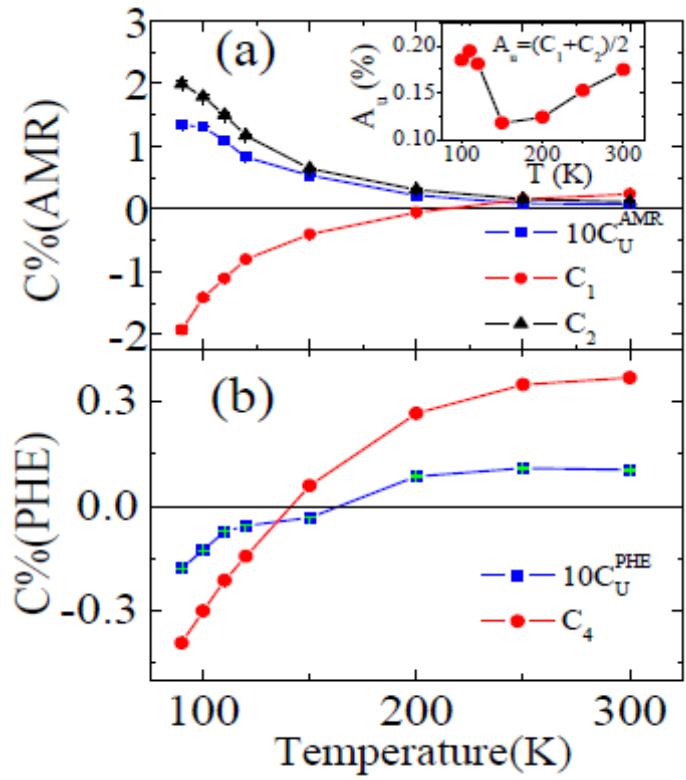

Fig.5. The fitted coefficients of (a) AMR with Eqn. (7) and (b) PHE with Eqn. (8) as a function of temperature. The inset in (a) shows the temperature dependence of the twofold terms $A_u$, resulting from the competition between $C_1$ and $C_2$.

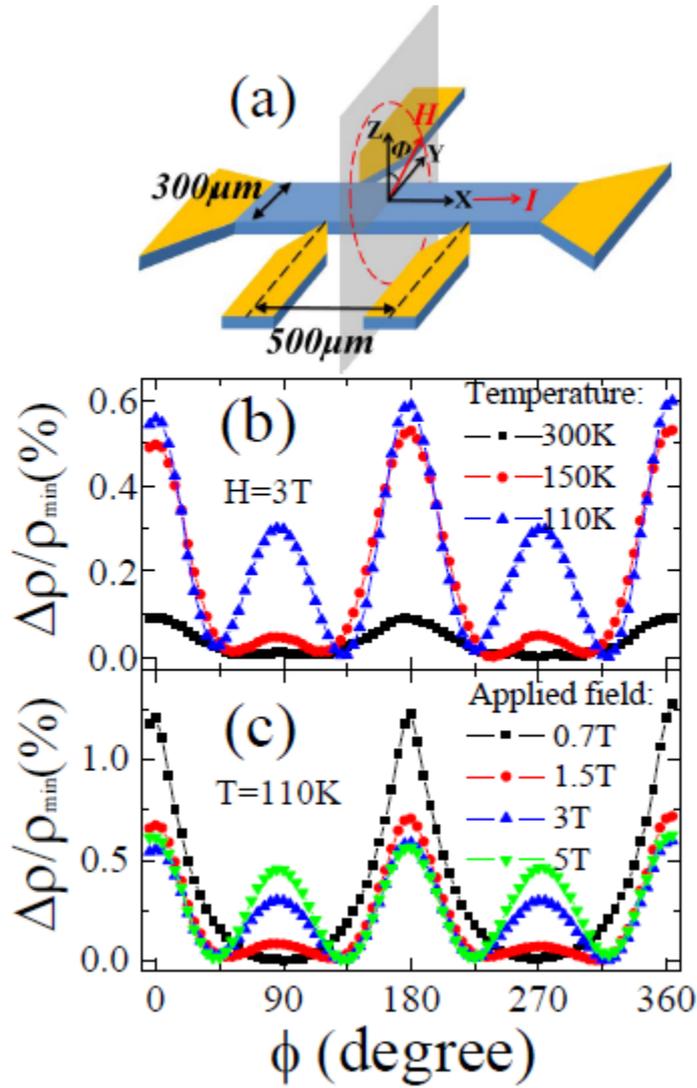

Fig.6. (a) Schematic drawing of the rotational plane perpendicular to the current. The angular dependent longitudinal resistivity of a 46.8nm $Fe_3O_4$ film (b) at different temperatures with a 3T applied field and (c) with different applied fields at T=110K.

*Corresponding author:wuyizheng@fudan.edu.cn